\documentclass{article}
            \usepackage{graphicx}
            \usepackage{amsmath,amssymb}
            \begin{document}
           \title{Relation between observers and effects of number valuation
           in science}
          \author{Paul Benioff\thanks{The author thanks Professor Ken Augustyn for the opportunity to present this work at a workshop on biological mentality. The opportunity to publish the work in a journal  with other workshop papers is appreciated.}\\
            Physics Division, Argonne National
           Laboratory,\\
           Argonne, IL 60439, USA \\
           email: pbenioff@anl.gov}
           \maketitle

            \begin{abstract}This paper is a small step towards the goal of constructing a coherent theory of physic and mathematics together.  It is based on two ideas, the localization of mathematical systems in space or space time, and the separation of the concepts of number from number value. The separation of number from number value along with the freedom of choice of number values at different points of space or space time enables the introduction of a space or space time dependent number valuation field.   The presence of a location dependent number value field affects theoretical descriptions of many physical and geometric quantities.  A simple geometric  example is worked out in detail, that of the length of a path.

            The localization of mathematical systems and the separation of number from number value or meaning both emphasize the role of observers.  The separation of number from number value shows the role of observers in that value or meaning are conscious observer related concepts.  Nothing, including numbers, has value or meaning to an unconscious observer. The localization of mathematical systems also shows the role of observers in that the mathematics that is potentially available to an observer is that at the same position as is the observer. This represents the mathematical knowledge that can reside in an observers brain.  As an observer moves in space or space time, the mathematical knowledge potentially available to the observer is the collection of mathematical systems at the same location as the observer. It is hoped that this work, which was begun in 2010, will lead to a better understanding of the relation between the foundations of mathematics and physics, and the role that observers play in this relation. \end{abstract}

           \section{Introduction}

           The relationship between the sciences and mathematics is a topic of some interest.  This can be seen by the literature responses \cite{Omnes,Plotnitsky} to Wigner's provocative paper on "The unreasonable effectiveness of mathematics in the natural sciences" \cite{Wigner}.  This work sparked the authors interest in the topic.  It led to the belief that there may be a coherent theory of mathematics and physics together \cite{BenCTPM,BenCTPM2}.  It is hoped that work to better understand the relation between mathematics, physics, and geometry will contribute to a better comprehension   of the  relationship between these three topics. Hopefully it  will lead to development of a coherent theory.

           This paper is a small step in this direction. Two concepts are introduced:  The existence of  mathematical systems of different types localized to points in space or space time, and the separation of two concepts that are usually conflated, that  of number from that of number value.

           The emphasis on the distinction between number and number value or meaning  emphasizes the relevance of  observers to mathematics.  The reason is that number value or meaning is an observer related concept. The value or meaning of numbers, is an essential part of consciousness.  Number values or meaning, as well as the values or meanings of many other elements, do not exist "out there" independent of an observer's consciousness.

           The existence of conscious observers is a necessary condition for numbers  to have values or meanings. This also implies that number value is a local concept in that it is localized within an observers brain.  The fact that observers at different locations all agree on the meaning and value of numbers and other mathematical elements is enabled by the use of  maps of mathematical systems between different locations.   These maps, and value changing maps, which are needed to describe theory predictions of physical and geometric quantities,  will be discussed later on.

           The idea of local mathematical systems is not new.  It is a part of the mathematics used by gauge theories in physics. The mathematical background for these theories consists of vector spaces at each point of space time \cite{Montvay}. The relations between vectors in these spaces are defined by unitary operators as gauge transformations.  The generators of the Lie algebra representations of these unitary operators correspond to the bosons (photons, $W$ and $Z$ particles,and gluons) of the standard model of physics.

           Here localization of mathematical systems is extended to numbers of different types and to mathematical systems of types that include numbers as part of their description. Vector spaces, operator algebras, group representations as matrices, are examples. In this work the main emphasis will be limited to numbers and vector spaces, both in physics and geometry.

           The other new element added here is the freedom of choice of number values at different locations in space, time or space time.  This idea has its origin in the Yang Mills \cite{YangM} freedom of choice of vectors in gauge theory.  This was the concept that the choice of which vector represents a proton (in isospin space) at one location does not determine the vector that represents a proton at another location.  This concept is applied to numbers in that the value of a number at one location does not determine its value at another location.

           The freedom of choice of number values is accounted for here by the introduction of a scalar valuation field, $g$.  This field takes real values, $g(x)$ at each location $x.$  This field assigns a scale or value factor to the local mathematical structures at each location.  It also is used in the definition of connections which give the relations between  mathematical structures and their components at different locations.

             An application of these concepts in geometry, will be given. This is followed by a description of the limitations that experiment places on the $g$ field.  A final section gives some details on the close connection between local mathematical structures and observers.

           To proceed  further it is necessary to show in detail the difference between number and number value.  This is done in the next section. This is followed by a description of mathematical systems of different types as structures  \cite{Shapiro,Barwise,Keisler} that satisfy a set of axioms relevant to the structure type under consideration.

           Historically the first use of number scaling or number valuation appeared in 2011 \cite{Ben1st}. This was followed by other work expanding the use of space and time dependent number scaling or valuation \cite{BenINTECH,BenFB,Ben2nd}.

           The scaling or valuation used in these papers and in this work is linear.  Recently this type of scaling has been generalized to a functional scaling by Czachor \cite{Czachor}.  This type of scaling was used to suggest a relation between generalized arithmetic and dark energy \cite{Czachor2}.  However, the possible space and time dependence of functional scaling was not described.

           \section{Number and number value}
           \subsection{Natural numbers}

           The simplest way to see the distinction between number and number value is with the natural numbers or nonnegative integers.
           Let\begin{equation}\label{nN} 0,\; 1,\; 2,\; 3,\; 4,\; \cdots=B_{N}\end{equation} represent a set of natural numbers. Here $0,1,2\cdots$ are the names of these natural numbers. They also denote the values of these numbers.  The values correspond to the position of each number in the implied well ordering in the set, $B_{N}.$

           The context in which the numbers in $B_{N}$ obtain their value is provided by a natural number structure, $\bar{N}$ where \begin{equation}\label{bN}\bar{N}=\{B_{N},+,\times,<,0,1\}.\end{equation} Here $+$ and $\times$ are addition and multiplication operations, $<$ is a well ordering relation, and $0$ and $1$ are constants. The structure $\bar{N}$ must satisfy the axioms for arithmetic.

           Consider the subset of natural numbers represented by \begin{equation}\label{BN2}0,\;\; 2,\;\; 4,\cdots=B_{N_{2}}.\end{equation}   These even numbers of $B_{N}$ can also be used to do arithmetic.  The values assigned to the numbers in $B_{N_{2}}$ correspond to their position in the well ordering inherited from that for $B_{N}.$. The natural number represented by $0$ has value $0$,  the number represented by $2$ has value $1$, the number represented by $4$ has value $2$ and so on.

           This shows explicitly that numbers and number values are distinct. The number that has value $2$ in $B_{N}$ is the number that has value $1$ in $B_{N_{2}}.$ In general the number that has value $2n$ in $B_{N}$ has value $n$ in $B_{N_{2}}.$

           One sees from this that numbers by themselves do not have value.  Value is provided by the context or  mathematical environment of the numbers, in this case, the well ordering.  The only exception to this is the number $0.$ Its value is independent of the context or well ordering of the set containing it.

            The values of numbers in $B_{N_{2}}$ can also be determined by the properties they have in a structure, $\bar{N}^{2}$ where \begin{equation}\label{bN2}\bar{N}^{2}=\{B_{N_{2}},+_{2},\times_{2},<_{2}, 0_{2},1_{2}\}.\end{equation} This structure can be regarded as providing the context or environment in which the numbers in $B_{N_{2}}$ obtain their value.  Note that $\bar{N}^{2}$ must also satisfy the axioms for arithmetic.

           The representation of numbers as values with scaling or value factor subscripts will be used throughout.  Thus $1_{2}$ is the number with value, $1,$ in $\bar{N}^{2}.$ In general, $a_{2}$ is the number with value $a$ in $\bar{N}^{2}.$

           There are two ways to relate the structures, $\bar{N}^{2}$ and $\bar{N}=\bar{N}^{1}.$ One is by a number changing, value preserving, map of $\bar{N}^{2}$ onto $\bar{N}^{1}.$  This is given by \begin{equation}\label{WN21}\begin{array}{c}a_{2}\rightarrow a_{1} \mbox{  for all numbers in $B_{N_{2}}$,} \\+_{2}\rightarrow +_{1},\;\;\times_{2}\rightarrow \times_{1},\\<_{2}\rightarrow <_{1}, 1_{2}\rightarrow 1_{1}.\end{array}\end{equation}The other method is by a number preserving, value changing map of $\bar{N}^{2}$ into $\bar{N}^{1}.$  This is given by \begin{equation}\label{ZN21}\begin{array}{c}a_{2}\rightarrow (\mbox{\Large$\frac{2}{1}$}a)_{1},\mbox{ for all numbers in $B_{N_{2}},$}\\ +_{2}\rightarrow +_{1},\;\; \times_{2}\rightarrow (\mbox{\Large$\frac{1}{2}$}\times)_{1},\\ <_{2}\rightarrow <_{1},\;\; 0_{2}\rightarrow 0_{1},\;\;1_{2}\rightarrow (\mbox{\Large$\frac{2}{1}$})1_{1}.\end{array}\end{equation}

           Note that $a_{2}$ is the same number in $\bar{N}^{2}$  as is $((2/1)a)_{1}=(2a)_{1}$ in $\bar{N}^{1}.$ The change in value reflects the change in context or environment in going from $\bar{N}^{2}$ to $\bar{N}^{1}.$ The number value, $1,$ is preserved in the denominator to be consistent with generalizations to be described.

           The structure with components  shown in Eq. \ref{ZN21} is given by \begin{equation}\label{bN21}\bar{N}^{2}_{1}=\{B_{N_{2}},+_{1}, (\frac{1}{2}\times)_{1},<_{1},0_{1},(\frac{2}{1}1)_{1}\}.\end{equation} The components of this structure are a representation of the components of $\bar{N}^{2}$ in terms of the components of $\bar{N}^{1}.$  Scaling of the multiplication operation is necessary to preserve arithmetic axiom validity for $\bar{N}^{2}_{1}.$

\begin{figure}[h!]\vspace{-4cm}{\includegraphics[scale=.5]
            {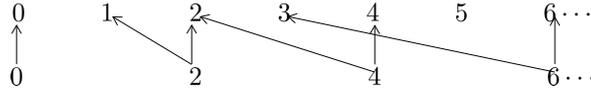}}\vspace{-2cm}\caption{Schematic representation of the actions of number changing, value preserving and number preserving value changing maps on $B_{N_{2}}.$
            The diagonal and vertical arrows show, respectively, the number changing value preserving and number preserving, value changing maps. The digits are names of numbers, either in $B_{N_{2}}$ or in $B_{N}.$} \label{RENV1} \end{figure}

           Figure \ref{RENV1} illustrates the actions of number changing value preserving and number preserving value changing operation on the numbers in $B_{N_{2}}.$  The vertical arrows represent number preserving value changing operations.  The diagonal lines  illustrate number changing value preserving operations. The digits denote numbers, either in $B_{N_{2}}$ or in $B_{N}.$

           The description of the two types of maps, applied to the even numbers, can be extended to the subset, $B_{N_{n}}$, of $B_{N}$ containing every $nth$ number.  The corresponding structure for these numbers is given by
           \begin{equation}\label{bNn}\bar{N}^{n}=\{B_{N_{n}},+_{n},\times_{n},<_{n},0_{n},1_{n}\}.\end{equation}Let $n$ be a non prime number with $m$ as a factor.  Then $B_{N_{n}}\subset B_{N_{m}}.$  The number preserving, value changing map  of the components of $\bar{N}_{n}$ to those of $\bar{N}_{m}$ can be represented by the structure, \begin{equation} \label{bNnm}\bar{N}^{n}_{m}=\{B_{N_{n}},+_{m},(\frac{m}{n}\times)_{m},<_{m},0_{m},(\frac{n}{m}1)_{m}\}.\end{equation}The scaling of the multiplication operation is necessary in order that $\bar{N}^{n}_{m}$ preserve the axioms of arithmetic.

           One sees from this that a number can have many different values.  If $n$ is a prime number then a number in $B_{N_{n}}$ has two representations as $a_{n}$ and $(na)_{1}$ If $n$ is not a prime number, then a number in $B_{N_{n}}$ can have several different representations.  As an example, the possible representations for the number that is the identity in $\bar{N}^{30}$ are given by  $$1_{30}=2_{15}=3_{10}=5_{6}=6_{5}=10_{3}=15_{2}=30_{1}.$$

           This description of the distinction between number and number value and the relations between natural number structures with different scaling or valuation factors has been given in detail. The reason is that natural numbers provide the simplest example of the effects, scaling or valuation factor change, have on number structures.

           The description for natural numbers applies with minor changes to integers. It also applies with more changes to rational,  real, and complex numbers. The main change for these number types is that the  base sets of numbers, $B_{Ra},$ $B_{R}$, and $B_{C}$ are unchanged under change of scaling or valuation factors. This is a consequence of the fact that these types of numbers  include division (or inversion)  as a basic operation.

           Structures for each of the three different types of numbers with different scaling factors are summarized below. Additional details are quite similar to those for the natural numbers.

           \subsection{Rational numbers}

           Let $t$ and $s$ be two nonzero rational scaling or valuation factors.  The corresponding  rational number structures for these two factors are\begin{equation}\label{Ras}\overline{Ra}^{s}= \{B_{Ra}, \pm_{s},\times_{s},\div_{s},<_{s},0,1_{s}\}\end{equation} and \begin{equation} \label{Rat} \overline{Ra}^{t}=\{B_{Ra}, \pm_{t},\times_{t},\div_{t},<_{t},0,1_{t}\}. \end{equation} If either $s$ or $t$ are negative,  the direction of the order relation needs to be reversed. There are no subscripts on $0$ as number and number value are the same for all scaling factors.

           The number values, $t$ and $s$ are referred to here as either scaling or value factors. The choice is left up to the reader.  The term, 'value factor', will be used in the discussions of the effect of  space or time dependence of these factors on the values of physical or geometric quantities.

           The structure that expresses the components of $\overline{Ra}^{s}$ in terms of those of $\overline{Ra}^{t}$ is given by a number preserving, value changing map of the components of $\overline{Ra}^{s}$ onto those of $\overline{Ra}^{t}.$ One obtains \begin{equation}\label{Rast}
           \overline{Ra}^{s}_{t}=\{B_{Ra},\pm_{t},\frac{t_{t}}{s_{t}}\times_{t},\frac{s_{t}}{t_{t}}\div_{t},
           <_{t},0,\frac{s_{t}}{t_{t}}1_{t}\}.\end{equation}In this equation $s_{t}$ and $t_{t}$ are rational numbers with values, $s$ and $t$ in $\overline{Ra}^{t}.$

           It is worth noting that for any pair, $s,t,$ of rational scaling factors, $(s_{t}/t_{t})1_{t}=((s/t)1)_{t}$ is the identity in $\overline{Ra}^{s}_{t}.$ This is seen from the fact that it satisfies the multiplicative identity axiom.  To see this let $(s_{t}/t_{t})a_{t}$ be any rational number in $\overline{Ra}^{s}_{t}.$  One has\footnote{The implied multiplications  and divisions are  $\times_{t}$ and $\div_{t}.$} $$\frac{s_{t}}{t_{t}}a_{t} (\frac{t_{t}}{s_{t}}\times_{t})\frac{s_{t}}{t_{t}}1_{t}= \frac{s_{t}}{t_{t}}a_{t} \leftrightarrow a_{t}\times_{t}1_{t}=a_{t}.$$ The implied multiplications and division in the terms of this equivalence are those in $\overline{Ra}^{t}.$ Note that the three structures in Eqs \ref{Ras}, \ref{Rat}, and \ref{Rast} are required to satisfy the rational number axioms.

            \subsection{Real and complex numbers}

           The description of the corresponding three structures for the real numbers, with $s,t$ real scaling factors,  is essentially the same as that for rational numbers.  Eqs. \ref{Ras}, \ref{Rat}, and \ref{Rast} also describe scaled real number structures provided the base set $B_{Ra}$ of rational numbers is replaced by the base set, $B_{R}$ of real numbers.  The real number structures must satisfy the axioms for real numbers.

           The structures for complex numbers are the same as those for the real and rational numbers provided the order relations, $<_{t}$ and $<_{s}$, are replaced by complex conjugation operations. Also $B_{R}$ is replaced by $B_{C},$ the base set of complex numbers.

           It may seem strange that the complex number $(s_{t}/t_{t})1_{t}$, with both $s_{t}$ and $t_{t}$ complex, is the identity and is a real number. This emphasizes the fact that the properties of complex numbers are determined by the environment provided by the structure containing them. The complex number, $(s_{t}/t_{t})1_{t}$, is real and is the identity only within the context or environment of the  structure, $\bar{C}^{s}_{t}.$

           \section{Application of number scaling to physics and geometry}

           The description of numbers and number scaling provides a good basis for determining the effects of number valuation on properties of physical and geometric systems.  The physical and geometric quantities affected are those described by integrals or derivatives over space, time, or space time.

           The mathematical arena for this description consists of a collection of structures for different types of mathematical systems at each point of space, time, or space time. The structure for each type of system at each location has a value or scale factor associated with it. (Technically this arena is well described by a fiber bundle.)

           The assumption that the value of a number at one location does not determine the value at another location is accounted for by the introduction of a real valued value field $g$. The value factor for mathematical systems at each location, $x,$ is given by $g(x).$ For example, real and complex number structures at $x$, with the $g$ field included are $\bar{R}^{g(x)}_{x}$ and $\bar{C}^{g(x)}_{x}.$ Scaled vector spaces\footnote{Scaling also affects vector spaces. The local vector space structure at $s$ is given by $$\bar{V}^{(g(x)}_{x}=\{B_{V},\pm_{g(x)},\circ_{g(x)},|-|_{g(x)},\psi_{g(x)}\}.$$  Here $\circ_{g(x)}$ denotes scalar vector multiplication, $|-|_{g(x)}$ denotes a vector norm, and $\psi_{g(x)}$ denotes an arbitrary vector in $\bar{V}^{g(x)}_{x}$ with value $\psi$. } of arbitrary dimensionality can also be  included as $\bar{V}^{g(x)}_{x}.$ The scalar field associated with $V^{g(x)}_{x}$ is $\bar{R}^{g(x)}_{x}$ if $\bar{V}^{g(x)}_{x}$ is a real space. The scalar field is $\bar{C}^{g(x)}_{x}$ if $\bar{V}^{g(x)}_{x}$ is a complex vector space.

\begin{figure}[h]\vspace{-3cm}{\includegraphics[scale=.5]
            {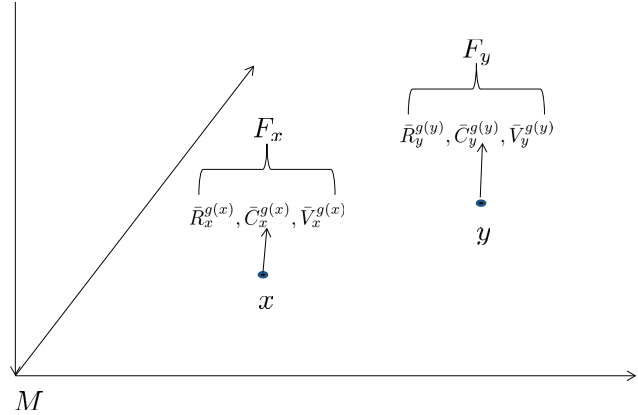}}\vspace{-1cm}\caption{Schematic showing local representations of real number, complex numbers and vector spaces at two locations, $x$ and $y$ in a three dimensional space, $M$.  The location dependent  value factors for the structures are $g(x)$ and $g(y)$ for the two locations. } \label{RENV3.1} \end{figure}

           Figure \ref{RENV3.1} shows an example collection of mathematical systems of different types at two points, $x$ and $y$ for the three dimensional space,  $M$.  The mathematical systems in the collection consist of structures for real numbers, complex numbers, and vector spaces.  The collection of systems at the points, $x$ and $y$ are denoted by $F_{x}$ and $F_{y}.$  The value factors for the points, $x$ and $y$ are denoted by $g(x)$ and $g(y).$

            There are many examples of physical quantities defined by integrals or derivatives over space.  For example, in quantum mechanics wave packets can be represented by integrals of the wave function over space.  Hamiltonians contain derivatives of the wave function. In geometry, lengths of paths are  integrals of infinitesimal path lengths along the path. These quantities are all affected by the presence of the $g$ field.

            The specific example that will be discussed here is the length of a path in $M$. It is simple and is illustrative of the effects of $g$ on physical and geometric quantities.

            Let $p$ be a path in $M.$  The path $p$ is parameterized by a real variable $t$. The path begins and ends at points, $p(0)=z$ and $p(1)=y.$ If $t$ represents time, then $p$  can be the description of the motion of a particle as it moves through space under the influence of a potential.

            The length of $p$ is given by the integral, \begin{equation}\label{Lp}
            L(p)=\int_{0}^{1}[\nabla_{t}p\cdot\nabla_{t}p]^{1/2}dt.\end{equation}Here
            \begin{equation}\label{natp}\nabla_{t}p\cdot\nabla_{t}p=\sum_{i=1}^{3}
            (\frac{dp_{i}}{dt})^{2}\end{equation}The $p_{i}$ are the path components in three directions in $M$.  The integrand of Eq. \ref{Lp} is a real number.  The quantity, $[\nabla_{t}p\cdot\nabla_{t}p]^{1/2}dt=dp(t)$ is the length of the path increment between $p(t)$ and $p(t+dt).$

            Eqs. \ref{Lp} and \ref{natp} give the path length in terms of the usual global number structures.  For these number and number value are the same.  For all $t$ values the integrand is a real number in just one real number structure.

            Here the mathematical arena is different.  One has local real number structures in which values of numbers are affected by the presence of a value field, $g.$ For each value of $t,$ the integrand of Eq. \ref{Lp} is a real number, $([\nabla_{t}p\cdot\nabla_{t}p]^{1/2})_{g(p(t))}$ in $\bar{R}^{g(p(t))}_{p(t)}$ at point $p(t)$ of the path.

            This representation of numbers is used throughout. Numbers are represented with subscripts.  The subscript denotes the value factor for the number structure containing  a number.  The quantity enclosed in parentheses is the value of the number in the structure. For example $([\nabla_{t}p\cdot\nabla_{t}p]^{1/2})_{g(p(t))}$ is the number in $\bar{R}^{g(p(t))}_{p(t)}$ that has number value, $[\nabla_{t}p\cdot\nabla_{t}p]^{1/2}.$

            The presence of the integrand in different real number structures for each $t$ value has the consequence that the path length integral, expressed by
            \begin{equation}\label{Lpg}L(p)=\int_{0}^{1}([\nabla_{t}p\cdot\nabla_{t}p]^{1/2})_{g(p(t))}dt,\end{equation}is not defined.  The implied summation of the definition of the integral is between real number structures at different locations in $M$.  Arithmetic operations, such as addition, are defined only within structures, not between them.

            This is fixed by parallel transforms of the integrands in the different structures to numbers in a single real number structure at an arbitrary reference location, $x$, in $M$.  The parallel transforms are determined by the values of $g$ at the points, $p(t)$ and $x.$

\begin{figure}[h]\vspace{-2cm}\hspace{-1cm}{\includegraphics[scale=.55]
            {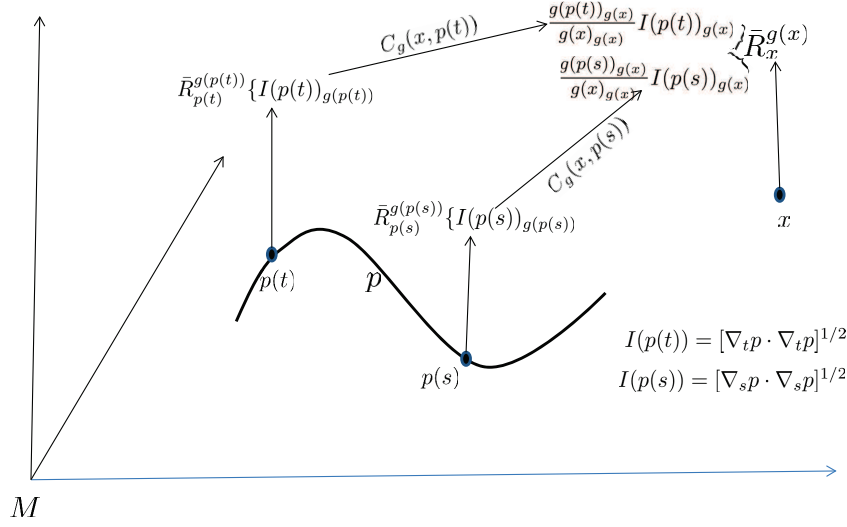}}\caption{Path length integrands for two path points, $p(t)$ and $p(s),$ are shown along with their membership in the real number structures, $\bar{R}^{g(p(t))}_{p(t)}$ and $\bar{R}^{g(p(s))}_{p(s)}$. Parallel transport of these integrands to the same numbers in $\bar{R}^{g(x)}_{x}$ is shown as being implemented by the connections, $C_{g}(x,p(t))$ and $C_{g}(x,p(s))$.  These are number preserving value changing maps.} \label{RENV3.2} \end{figure}

            Figure \ref{RENV3.2} illustrates the  values of the path length integrand in the real number structures at two path points, $p(t)$ and $p(s).$ The membership of these integrands in $\bar{R}^{g(p(t))}_{p(t)}$ and $\bar{R}^{g(p(s))}_{p(s)}$ is indicated by the curly brackets. The effect of parallel transport of these integrands to numbers in $\bar{R}^{g(x)}_{x}$ is indicated by the connections, $C_{g}(x,p(t))$ and $C_{g}(x,p(s)).$

            For each value of $t$ the transformed integrand is given by \begin{equation}\label{gptx}
            (\frac{g(p(t))}{g(x)})_{g(x)}([\nabla_{t}p\cdot\nabla_{t}p]^{1/2})_{g(x)}=(\frac{g(p(t))}{g(x)} [\nabla_{t}p\cdot\nabla_{t}p]^{1/2})_{g(x)}.\end{equation}The path length at reference point, $x$ is given by \begin{equation}\label{Lpx}L(p)_{x,g(x)}=(\frac{1}{g(x)})_{g(x)}\int_{0}^{1}(g(p(t) [\nabla_{t}p\cdot\nabla_{t}p]^{1/2})_{g(x)}dt_{g(x)}.\end{equation}

            This equation expresses the path length as a number in $\bar{R}^{g(x)}_{x}.$  The value of the number, as a length value, is obtained by removal of the subscript, $g(x)$ everywhere in Eq. \ref{Lpx}.

            The presence of the value field has several consequences for the path length that are not present in the usual case with $g(x)=1$ everywhere.\footnote{This condition can be relaxed to $g(x)$ is a constant, $c$ everywhere.} For example the path length is different for paths that are space translations or rotations of $p$.   If $q$ is the translation or rotation of $p,$  the length of the translated path, $L(q)_{x}$ is given by\begin{equation}\label{Lqx}L(q)_{x,g(x)}= (\frac{1}{g(x)})_{g(x)} \int_{0}^{1}(g(q(t))[\nabla_{t}q\cdot\nabla_{t}q]^{1/2})_{g(x)}dt_{g(x)}.\end{equation}Differences in $L(p)_{x,g(x)}$ and $L(q)_{x,g(x)}$ result from differences in the values of the $g$ field along points of $q$ compared to those along points of $p.$

            The effect of the value field also shows up in  changes in the location of the reference point in $M$. The path length, as a number in $\bar{R}^{g(y)}_{y}$, is given by \begin{equation}\label{Lpy}L(p)_{y ,g(y)}=(\frac{1}{g(y)})_{g(y)}\int_{0}^{1}(g(p(t)[\nabla_{t}p\cdot\nabla_{t}p]^{1/2})_{g(y)}dt_{g(y)}. \end{equation} This expression is the same as that for the length in Eq. \ref{Lpx} with $y$ replacing $x$ everywhere.

            The only difference between the number values of $L(p)_{x,g(x)}$ and $L(p)_{y,g(y)}$ comes from the replacement of $(1/g(x))_{g(x)}$ in Eq. \ref{Lpx} with $(1/g(y))_{g(y)}$ in Eq. \ref{Lpy}.  The difference in the number values of the lengths can be expressed as \begin{equation}\label{Lpyx}
            L(p)_{y}=\frac{g(x)}{g(y)}L(p)_{x}.\end{equation}

            The description of the effect of local number structures and the presence of the value field on path lengths extends to other quantities. Included are wave packets as space integrals of wave functions, derivatives in Hamiltonians, and many other quantities \cite{BenINTECH,BenFB,Ben2nd}.  These will not be discussed here as they add nothing to  the purview of this work.

            \section{Restrictions on the value field, $g$}

            To date there is no experimental evidence for the presence of the $g$ field.  This implies that the deviation of the $g$ field values  from unity must have been too small to have been detected.  The deviations need not be zero.  However they must be so small that they are less than the statistical uncertainties of experiments done so far.

            The restrictions on the $g$ field are based both on experimental non detection of the field presence and on the fact that all theory computations and experiments are done by us.  Locations of theory computations and experiments are limited to space and time regions  in the universe that are occupied or can be occupied by us. Experiments or computations in regions of space and time that are not occupiable by us cannot be done by us.

            It follows that the deviations of the $g$ field from unity must be within experimental uncertainties in all regions of cosmological space and time that are occupied by us.  We, as observers, are the ones doing the experiments to test theories.  We are the ones making the computations of theoretical predictions to test with experiment.

            This restriction on $g$ is too weak.  As stated it applies to regions on the surface of the earth that are occupied by us.  However, it should be extended to regions of space and time that are occupiable by us, now or in the future. This greatly extends the region to include much of the space and time in and around the solar system. The international space station is an example of  the extension of the region beyond the earth.

            This extension is based on the unproven assumption that no experiment done by us, now or in the future, will reveal the presence of variations in the $g$ field at the location of the experiment.  This is a conservative assumption. It may well be that in the future an experiment will be done that will reveal the presence of the value field. Until this happens it is best to assume that outcomes of experiments, done by us at different space and time locations will not  ever show a space and time dependence attributable to local variations in the $g$ field.

             This restriction is still too weak.  It should also be extended to locations of other intelligent beings on other planets. The possibility that these beings will do or have done  experiments that show variations in experiment outcomes that are due to variations in the $g$ field in their local region should be excluded.

              This a very conservative and extreme restriction.  Here it is relaxed by the limitation that variations in the $g$ field that are detectable by experiment  should be limited to regions of space and time to regions occupiable by us or by intelligent beings on other planets  at locations that are within effective two way communication with us.

            A literature estimate of the size of this  region is a sphere of radius about $1200$ light years centered on us \cite{AmSci}. The specific size of the region is not important.   Here the only requirement on the restricted region is that it is a small fraction of the cosmological universe. Outside this region there are no restrictions on $g$. The field can fluctuate rapidly, or slowly, or not at all.

            There are several physical fields that exist or have been proposed that satisfy the restriction and are scalar fields. These include the inflaton \cite{Infl} and quintessence \cite{Quint} and others \cite{Rinaldi}. If one includes the possibility that the $g$ field has already been detected but in a different guise, then the scalar Higg's field \cite{Higgs} becomes a candidate. One should also include dark matter \cite{DM} and dark energy \cite{DE} as possible candidates.

             It may be that the $g$ field is none of these candidate fields.  It is possible that the $g$ field, as a value field, has some connection with the close relation between the foundations of mathematics and physics.  Since number value is a observer related concept, it also brings observers into the picture.

            \section{The role of the observer}
            It should be clear from the previous sections that the observer plays an important role in the interpretation and meaning of mathematics and physics and the relation between the two subjects. This is seen by the clear separation between the concepts of number and number value, both in mathematics and in physics. Numbers, either in physics or mathematics, have meaning or value only within the mind of a conscious observer. Of course numbers are not special in this regard.   The meaning or value of things in general, exists only within an observers mind. Absent all observers, nothing has meaning or value, including this sentence.

            The mathematical arena of local mathematical structures and their use in describing physical quantities brings  the observer into the picture.  A good way to see this is  by the use of fiber bundles in mathematics.  For the purposes of this work, it is sufficient to note that a fiber bundle consists of fibers located at each point of space or space time.  A fiber, $F_{x}$ at the location, $x$ contains mathematical structures of  different types.  These can include structures for each type of number, vector spaces, algebras, and other mathematical systems that include numbers in their axiomatic description. In actual use the fiber contents are tailored to the problem to be solved.

            Each observer, including the readers and author of this paper, are physical systems of great complexity.  The motion of each of us traces out a path $p$ in space  parameterized by time.  The position of an observer at time $t$ is denoted by $p(t).$

            Here the basic assumption underlying the relation between an observer and the local mathematical systems at different locations is that the mathematics that has meaning to an observer is limited to that at the location of the observer. Mathematical structures at locations that are different from that of an observer, are not accessible to the observer unless he or she moves to that location.

            It follows that meaningful mathematics, including that used to make theory predictions in physics, is contained in the fiber $F_{p(t)}$ at the location, $p(t),$ of the observer.  This is the mathematics and physics that has meaning or value to the observer at $p(t).$  If there are $n$ observers, $O_{j}$ where $j=1,2,\cdots,n,$  each moving on paths, $p_{j}$, then the mathematics and physics that has meaning  or value to  the observer, $O_{j},$ at path point $p_{j}(t)$ is that contained in the fiber, $F_{p_{j}(t)}.$

            The meaning or values of the  numbers  in the structures in $F_{p_{j}(t)},$ to the observer at $p_{j}(t),$ are determined by the value field, $g.$  The $g$ field provides a value or meaning to the numbers at each space location. This includes each point of each  path. The $g$ field provides meaning or value to elements such as numbers of vectors at each path location. For an observer $O_{j}$ at location, $p_{j}(t)$, the values of real numbers  are determined by the real number structure, $\bar{R}^{g(p_{j}(t))}_{p_{j}(t)}.$ The meaning or values of properties of vectors to the observer, $O_{j}$ at $p_{j}(t)$ are determined by the vector structure $\bar{V}^{g(p_{j}(t))}_{p_{j}(t)}$ and its associated scalar structure.

\begin{figure}[h]\vspace{-1cm}{\includegraphics[scale=.5]
            {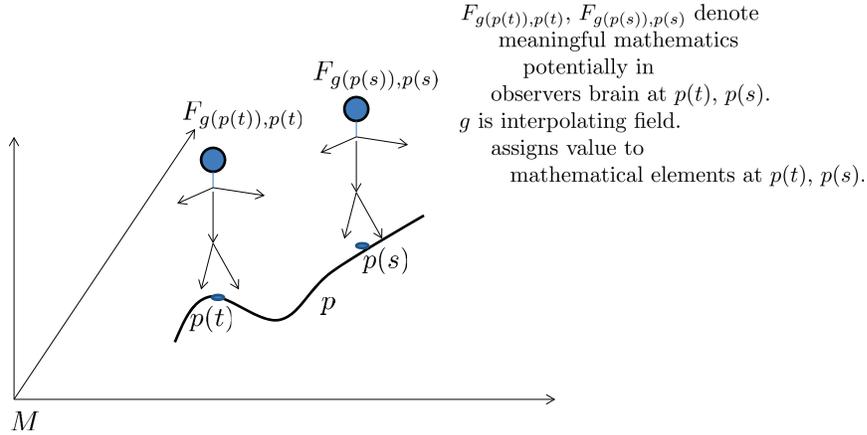}}\vspace{-2cm}\caption{Illustration of an observer on a path, $p,$ in three dimensional space, $M.$ The figure shows the observer at two path  points, $p(s)$ and $p(t),$  at two times, $s$ and $t.$ The  two collections of local mathematical structures available to the observer at these two times are denoted by $F_{g(p(t)),p(t)}$ and $F_{g(p(s),p(s)}.$  The value factors for the structures in the collections are shown as subscripts, $g(p(t))$ and $g(p(s)).$ The structures in the two collections are part of the potential knowledege of mathematics an observer can have at the two points.} \label{RENV5.1} \end{figure}

            Figure \ref{RENV5.1} provides an illustration of the relation between an observer at two different times on a path $p$ and the  local mathematics that has value to the observer at the two times. The denotation of the fibers in the figure includes the values of $g$  at the path locations.

            The value factor $g(p_{j}(t))$ provides an interpretation or value to the real numbers and vector properties for the observer, $O_{j}$ at $p_{j}(t).$ In acts like a space dependent decoding function in that it provides meaning or value to numbers or vector properties at each space  point.

            So far the $g$ field has been described as a space dependent function.  However it can easily be extended to depend also on time. This is the case in special relativity where the paths of observers in space time are world lines.  If $p_{j}$ denotes the world line of observer, $O_{j},$ the  space and time location of the observer at proper time, $\tau$ is given by   $p_{j}(\tau).$ The mathematics available to $O_{j}$ at $p(\tau)$, is contained in the fiber, $F_{p_{j}(\tau)}.$ In this case the $g$ field depends on both space and time where $g(x,t)$ is the value of $g$ at $x,t.$\footnote{The notation usually used in special relativity replaces the three dimensional $x$ and $t$ with the four dimensional $x=x_{\mu}$, where $\mu =0,1,2,3.$}

            The fact that the mathematics used by an observer  to make theoretical predications is local to the observers location has nothing to do with the physical or geometric quantities  described by the local mathematics.  The descriptions can refer to  properties of very far away systems such as galaxies, black holes, and other cosmological quantities.  The  local mathematics can also describe very small systems, such as atoms, nuclei, and elementary particles.

            It must be emphasized that the local mathematics of numbers, vectors and their values at $x$ will appear  the same to an observer, $O_{x}$ at $x$ as will the local mathematics of numbers, vectors and their values  at $y$ appear to an observer, $O_{y}$, at $y.$ This is the case even if the values of the $g$ field are different at the two locations. The observers at these locations will be in complete agreement on the properties of numbers, vectors and their values.

            For example, if $O_{y}$ sends the message  "$3.76$ is  a rational number with value $3.76$" to $O_{x}$ then $O_{x}$ will respond, "I agree". In this example $3.76$ is a string of symbols  at $O_{y}'s$ location. To $O_{y}$ it denotes the number with value $3.76.$ Transport of the symbol string, or equivalently, the information contained in the symbol string to location, $x,$ gives the same number at the location of $O_{x}.$  In the presence of the $g$ field,  the same numbers at $x$ will have a different value, $(g(y)/g(x)3.76.$ However the difference between the value, $3.76$ and the value $g(y)/g(x))3.76,$ will not be apparent to $O_{x}$ because the restrictions on the $g$ field limit the ratio, $g(y)/g(x)=1+\delta,$ where $\delta$ is too small to  be observable or to affect an observers awareness of the number. Note that this restriction applies only to the local region of the universe as described in the last section.

            As another example, both observers will agree with the statement "The length of path $p,$ given by
            \begin{equation}\label{Lpgz}L(p)_{g(z)}=(1/g(z))_{g(z)}\int_{0}^{1}(g(p(t))[\nabla_{t}p\cdot \nabla_{t}p]^{1/2})_{g(z)}dt,\end{equation}at reference location $z$, is a number with value $67.4469$."  $L(p)_{g(z)}$ denotes  the same number to $O_{y}$ as it does to $O_{x}$. However, the $g$ field restrictions imply that  the difference in the values  of the numbers at the two locations, $y$, and $x,$ are too small to be apparent to the observers.

            The agreement between observers is  also a consequence of the fact that the validity of the axiom sets for the different types of mathematical systems holds for structures at all space locations and for all $g$ values.  This requirement is essential.  Without it the whole setup in this paper would make no sense.

            \section{Summary and conclusion}

            In this paper some consequences of the separation of the concepts of number and number value have been examined.  The resulting freedom of choice of values of numbers  at different locations was accounted for by use of a scalar valuation field, $g$ that depends on space location. For each space location, $x$, the real number value, $g(x),$ was the scaling or valuation factor for number structures of different types. A real number, by itself has no value.  It acquires value or meaning only as a member of a real number structure, such as $\bar{R}^{g(x)}_{x}.$ The value of $g(x)$ along with the properties of the number in $\bar{R}^{g(x)}_{x},$ determine the value of the number at location $x$.  If $g(x)$ varies with  location, then the  value of the number is location dependent.

            As would be expected  the location dependence of $g$ affects the values of many physical and geometric properties.   This is the case for properties defined by integrals or derivatives over space. This was seen in the specific example of the length of a path in geometry. The effect is shown in the presence of  path position dependent $g$ values in the integrand of Eq. \ref{Lpx}. The effect on wave packets as space integrals of wave functions and on derivatives in Hammiltonians was mentioned.

            The lack of direct experimental evidence for the presence of the $g$ field was seen to place restrictions on the space and possible time dependence of $g$. In essence the restriction is that variations in $g$ must be below experimental error in all regions of the universe that are occupiable either by us or by intelligent beings that are within effective two way communication distance from us.  Here the size of this restricted region, estimated to be a sphere of about $1200$ lightyears in radius \cite{AmSci}, should be a small fraction of the size of the universe.

            Some candidate scalar fields  that satisfy this condition were noted.  These include the inflaton and quintessence.  Dark energy and dark matter also seem to satisfy this restriction.  The scalar Higg's field was also noted as a possible candidate.

            The description of the effects of space and time dependent number valuation on geometric and physical quantities was limited to three dimensional space.  This was done to emphasize the basic points of this work:  the distinction between number and number value and the effects of a space dependent value field.  Extension to descriptions in the space time of special relativity are straightforward \cite{BenFB}.

            Finally and most important, the role of the observer in this work must be emphasized.  The already appears in the separation of the concepts of number from number meaning or value. Number  value is an observer related concept. Number values are elements of an observers consciousness.  They reside in the brain in some form.

            The use of local mathematical structures at each space point also reflects the fact that the mathematical knowledge in the mind of an observer is local. It resides within his or her brain. As an observer moves along a path in space, the mathematical knowledge that is potentially available  is that at the observer's location.

            As might be surmised there is much work to do in further exploration of the use of local mathematical structures and a space and time dependent value field. One assumption that needs change is the  location of observers at points in space and time.  This is a fiction.  As large physical systems, observers occupy regions of space and time, not points.  In addition the large amount of information in an observers brain requires a minimum amount of space. One may hope that the value field will turn out to bind mathematics and physics more closely together, with observers playing an essential role in the process.

            \section*{Acknowledgement}
            This material is based upon work supported by the U.S. Department of Energy, Office of Science, Office of Nuclear Physics, under contract number DE-AC02-06CH1135

           \end{document}